\documentclass[final,3p,times]{elsarticle}
\usepackage{amssymb}
\usepackage{amsmath}
\biboptions{sort&compress} 
\usepackage{hyperref}
\journal{Journal of Subatomic Particles and Cosmology}
\begin{document}
\begin{frontmatter}
\title{Light hadron production measurements with Au+Au Collisions from $\sqrt{s_{NN}} = 3.2$--$4.5$ GeV with STAR}
\author[Davis]{Mathias C. Labonté (\textit{for the STAR collaboration})\corref{cor1}}
\ead{mlabonte@ucdavis.edu}
\cortext[cor1]{Corresponding author}
\affiliation[Davis]{organization={University of California, Davis},
             addressline={One Shields Avenue},
             city={Davis},
             postcode={95616},
             state={California},
             country={USA}}
\begin{abstract}
One of the main physics goals of the Beam Energy Scan program at RHIC is to study the QCD
phase diagram, specifically around the phase transition between the quark-gluon plasma and
hadronic matter. Beam Energy Scan Phase-I studied Au+Au collisions from center-of-mass energy ($\sqrt{s_{NN}}$) of 7.7 to
62.4 GeV. Beam Energy Scan Phase-II extended these measurements in several important ways, one of which was
the addition of a fixed-target program that pushed the collision energy down to 3.0 GeV (or baryon
chemical potential, $\mu_B$, up to 720 MeV). Fixed-target collisions at STAR allow for a more extensive
scanning of the QCD phase diagram to an important region where the QCD critical point may lie, and
to a region dominated by dense baryonic matter. One key measurement in the fixed-target program is
the spectrum of the lightest hadrons ($\pi^{\pm}$, $K^{\pm}$, p) as a function of transverse momentum, rapidity,
and collision centrality. From the $p_T$ spectra, a blast-wave model is used to study the temperature at kinetic freeze-out and the surface velocity of the expanding matter. These results provide important input to models of heavy ion collisions at these energies, and can help constrain the equation of state of QCD matter. Here, results are shown for four collision energies in the fixed-target range: $\sqrt{s_{NN}} = 3.2$, 3.5, 3.9, and 4.5 GeV.
\end{abstract}
\end{frontmatter}
\section{The STAR Beam-Energy Scan}
The Beam Energy Scan (BES) program at the Relativistic Heavy Ion Collider (RHIC) aims to explore the QCD phase diagram, particularly the region of high baryon chemical potential ($\mu_B$). At vanishing $\mu_B$, lattice QCD calculations indicate a smooth crossover between the Quark-Gluon Plasma (QGP) and hadronic phases~\cite{Aoki:lattice}. At larger $\mu_B$, model calculations suggest a first-order phase transition line terminating in a critical point~\cite{Stephanov:critical}, although the existence and location of such a feature remain experimentally unconfirmed. Mapping the QCD phase diagram in the high-$\mu_B$ region is therefore an important physics goal of the BES program. BES Phase-I studied Au+Au collisions in collider mode at $\sqrt{s_{NN}}$ from 7.7 to 62.4 GeV, where mid-rapidity kinetic and chemical freeze-out parameters were extracted, thereby mapping the chemical freeze-out line and characterizing kinetic freeze-out up to $\mu_B \approx 420$~MeV~\cite{STAR:BESI:freezeout}.
BES Phase-II extended the program by introducing a gold fixed target, allowing STAR to study collisions at lower center-of-mass energies (corresponding to higher $\mu_B$) in addition to those accessible in collider mode. Data have been collected at $\sqrt{s_{NN}} = 3.0 - 13.7$ GeV in fixed-target mode (FXT).

The FXT configuration introduces several experimental challenges; the acceptance and location of mid-rapidity change with collision energy, with mid-rapidity lying outside the nominal STAR acceptance at $\sqrt{s_{NN}} = 7.7$ GeV. The endcap Time-of-Flight (eTOF) and inner Time Projection Chamber (iTPC) upgrades extend STAR's coverage and help recover access to mid-rapidity for several FXT energies~\cite{STAR:eTOFPerf,STAR:iTPC}. In these proceedings, the bulk properties of the system created in fixed target collisions at $\sqrt{s_{NN}} = 3.2$, 3.5, 3.9, and 4.5 GeV are explored by fitting the $p_T$ spectra of charged pions, charged kaons, and protons to a blast-wave model \cite{SchnedermannSollfrankHeinz}. This allows for the extraction of the temperature at kinetic freeze-out, and the surface velocity of the expanding medium. Then, proton $dN/dy$ are studied in order to quantify the amount of baryon stopping observed, and a systematic scan is performed across energy and centrality (0-5\%,~5-10\%,~10-20\%).
\section{Experimental methods}
The Time Projection Chamber (TPC) provides particle identifcation (PID) for low momentum particles by measuring energy loss ($dE/dx$). The TOF extends PID capabilities to higher momenta, where particle species are no longer identifiable by $dE/dx$ alone. Identified particle yields are then corrected for detector efficiency, finite acceptance of the experiment, and energy loss as they traverse the experimental material. A feed-down correction from weak decays (\textit{e.g.} $\Lambda \rightarrow \pi^- + \rm p$) has not been applied here. The contribution of feed-down protons to the inclusive yields  at midrapidity is roughly 5\% at 3.2 GeV , and grows to approximately 20\% at 4.5 GeV. Proton yields that are corrected for feed-down contributions will be provided in forthcoming publications.
\section{Results}
\begin{figure*}[!h]
\centering
  \includegraphics[width=0.85\textwidth]{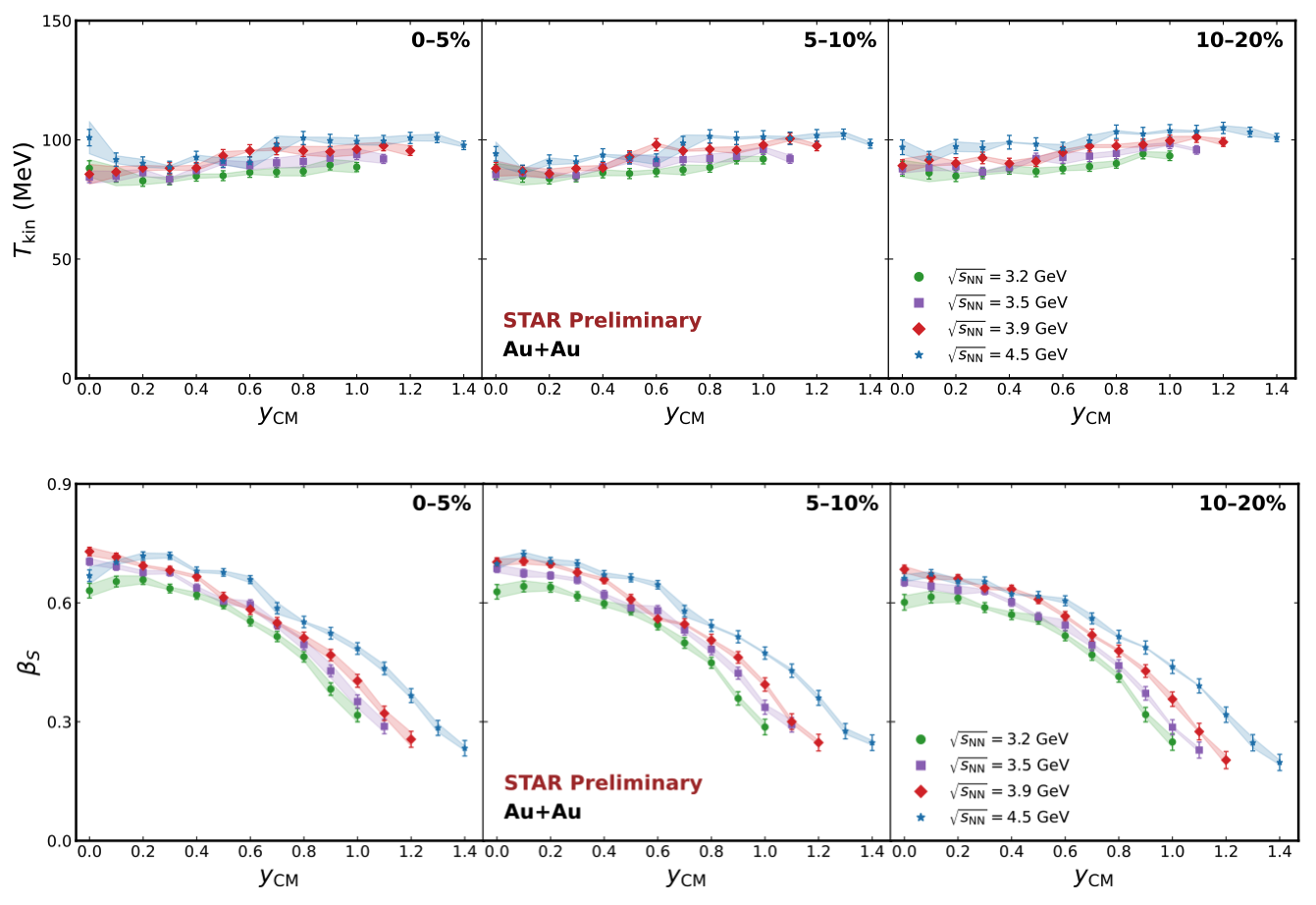}
  \caption{Upper Panel: Temperature at kinetic freeze-out extracted via the blast-wave model.  Lower Panel: Surface velocity extracted via the blast-wave model. Results correspond  to Au+Au collisions at $\sqrt{s_{NN}} = $3.2 (green circles), 3.5 (purple squares), 3.9 (red diamonds), and 4.5 (blue stars) GeV and 0-5\%, 5-10\%, 10-20\% most central collisions. Shaded bands correspond to systematic error, and vertical bars represent statistical error.}
  \label{fig:bulk}
\end{figure*}
Transverse momentum ($p_T$) spectra of $\pi^{\pm}$, $K^{\pm}$, and $p$ are measured as a function of rapidity, centrality, and collision energy. The spectra for all particle species are fitted simultaneously using a blast-wave function to extract the kinetic freeze-out temperature $T_{\mathrm{kin}}$ and the surface velocity $\beta_s$. The blast-wave model assumes a thermalized source and a transverse flow velocity profile $\beta_T(r) = \beta_s (r/R)^n$~\cite{SchnedermannSollfrankHeinz}, with the profile exponent is fixed to $n=1$ in this analysis. The $\pi^{\pm}$ spectra are fitted for $p_T > 0.3$ GeV/$c$, where the contribution from resonance decays is reduced and the spectra are better described by the blast-wave model.
\begin{figure*}[!h]
\centering
  \includegraphics[width=0.46\textwidth]{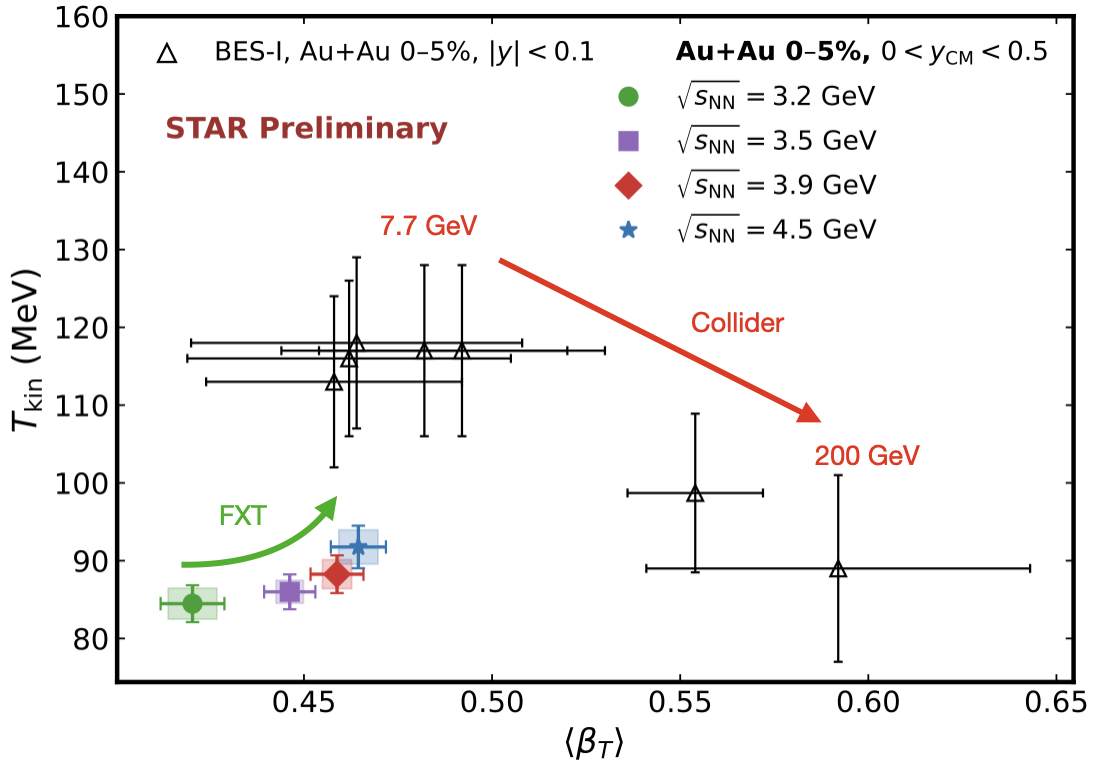}
  \caption{STAR FXT $T_{\rm kin}$ and $\langle \beta_T \rangle$ measurements averaged over rapidity ($0 < y_{\rm cm} < 0.5$) and compared to BES-I results (open markers) \cite{STAR:BESI:freezeout}.  BES-II results correspond  to Au+Au collisions at $\sqrt{s_{NN}} = $3.2 (green circles), 3.5 (purple squares), 3.9 (red diamonds), and 4.5 (blue stars) GeV, 0-5\% most central collisions. Shaded bands correspond to systematic error, and vertical bars represent statistical error.}
  \label{fig:world}
\end{figure*}

The extracted $T_{\mathrm{kin}}$ and $\beta_s$ are shown in Fig.~\ref{fig:bulk} as a function of rapidity in the center-of-mass frame ($y_{\mathrm{cm}}$) for three centrality bins (0-5\%, 5-10\%, 10-20\%) and four collision energies ($\sqrt{s_{NN}} = 3.2$, 3.5, 3.9, and 4.5 GeV). Across the FXT energy range, $T_{\mathrm{kin}}$ is largely independent of rapidity, with only a weak dependence on collision energy. The transverse flow velocity $\beta_s$ depends strongly on rapidity, and only has a weak dependence on energy. Figure~\ref{fig:world} shows the most central STAR $T_{\rm kin}$ and $\langle \beta_T \rangle$ values averaged over $0 < y_{\rm cm} < 0.5$, and compared with BES-I results \cite{STAR:BESI:freezeout}. As the collision energy increases, the results trend toward the values measured in collider-mode Au+Au collisions at $\sqrt{s_{NN}} = 7.7$ GeV.

\begin{figure*}[!h]
\centering
  \includegraphics[width=0.8\textwidth]{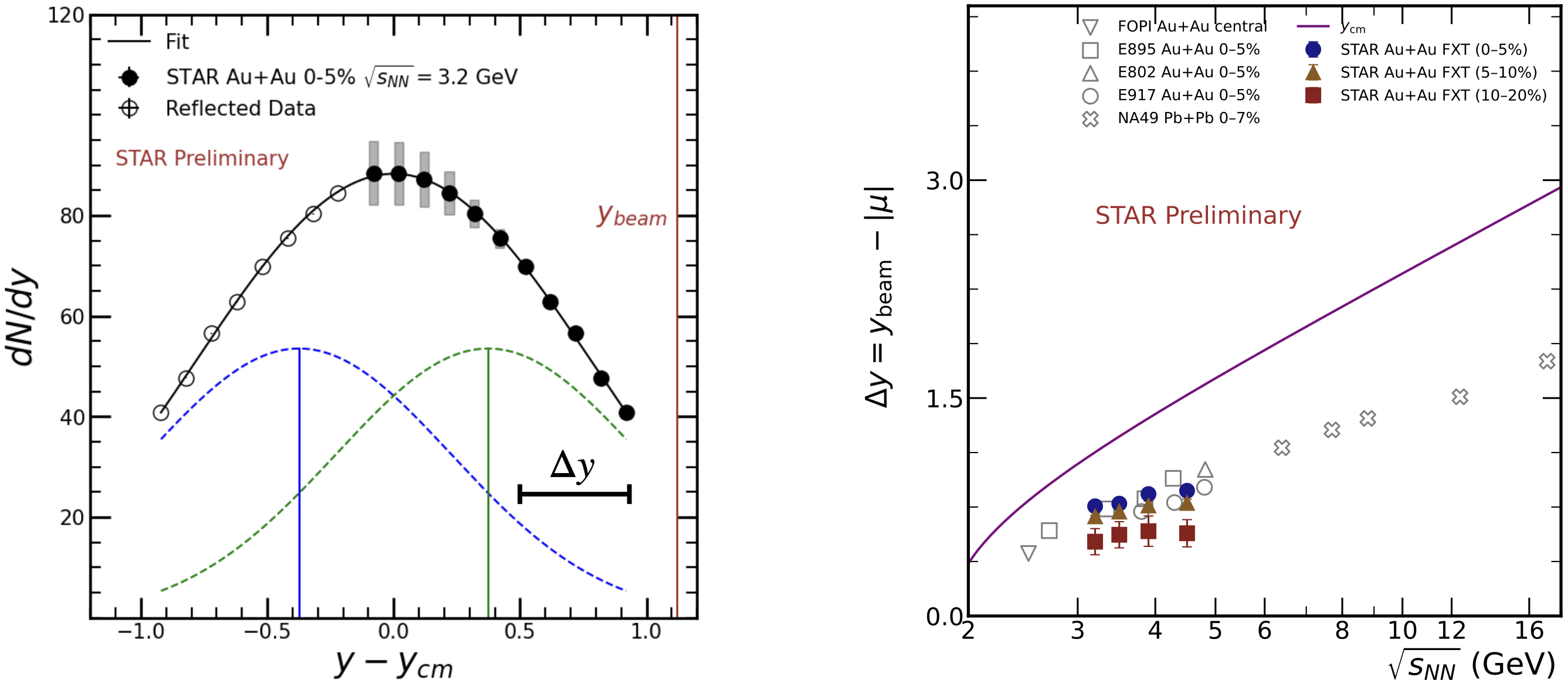}
  \caption{Left: $dN/dy$ of protons at $\sqrt{s_{NN}} = $3.2 GeV. Closed points are measured data, and open markers represent reflected data around $y_{\rm cm}$ = 0. Dashed lines indicate the double gaussian fit to the data modelling the projectile and target participants. Vertical solid lines represent the central value of the Gaussian. Right: Systematic scan of the stopping parameter $\Delta y$ as a function of $\sqrt{s_{NN}}$ measured by the STAR experiment for 0-5\% (blue circles), 5-10\% (brown triangles), and 10-20\% (red squares). Results are drawn alongside the most central available world data in the open markers \cite{FOPI:stopping, E895:stopping, E802:protons, E917:stopping, NA49:stopping, NA49:stopping2, NA49:stopping3}.}
  \label{fig:stopping}
\end{figure*}

The $p_T$-integrated yields ($dN/dy$) are obtained by extrapolating the measured spectra to $p_T = 0$ using the blast-wave model. The proton rapidity distribution, shown in the left panel of Fig.~\ref{fig:stopping}, is fitted with two symmetric Gaussians with mean~$\pm \mu$ and width~$
\sigma$ to model the stopped projectile and target participants. The baryon stopping observable is then defined as $\Delta y = y_{\mathrm{beam}} - |\mu
|$. The extracted $\Delta y$ values are shown as a function of $\sqrt{s_{NN}}$ in the right panel of Fig.~\ref{fig:stopping}, together with measurements from FOPI~\cite{FOPI:stopping}, E895~\cite{E895:stopping}, E802~\cite{E802:protons}, E917~\cite{E917:stopping}, and NA49~\cite{NA49:stopping, NA49:stopping2, NA49:stopping3}. 
The STAR results exhibit a smooth energy dependence, following the trend established by the E895 measurements at $\sqrt{s_{NN}} = 3.2$ and 3.9 GeV, while the result at $\sqrt{s_{NN}} = 4.5$ GeV lies between the E895 and E917 measurements. For the most central collisions, $\Delta y$  is roughly two thirds of beam rapidity (shown as a purple line) across all energies displayed. Less stopping is observed in less central collisions. This centrality ordering is the result of the smaller number of binary nucleon-nucleon interactions in peripheral collisions, which reduce the longitudinal momentum loss per participant nucleon. A dip in the baryon stopping trend as a function of collision energy has been associated with a softening of the equation of state and a possible first-order phase transition~\cite{Ivanov:stopping}. More precise measurements at higher FXT energies are needed to determine whether such a feature is present.
\section{Conclusion}
A blast-wave analysis of the bulk properties of the system produced in STAR fixed-target Au+Au collisions has been presented at $\sqrt{s_{NN}} =$ 3.2, 3.5, 3.9 and 4.5 GeV. $T_{\mathrm{kin}}$ shows only a weak dependence on rapidity and collision energy across the FXT range, while $\beta_s$ falls off away from mid-rapidity and grows slowly with increasing collision energy. The extracted baryon stopping parameter $\Delta y$ follows the trend established by AGS measurements, with reduced stopping in less central collisions consistent with a smaller fraction of participant nucleons undergoing multiple binary interactions. These measurements provide essential input for transport models of heavy ion collisions in this energy range, and can constrain the equation of state of dense baryonic matter probed by the FXT program. Future work will focus on extracting chemical freeze-out parameters using thermal models (Thermal-FIST~\cite{ThermalFIST}, THERMUS~\cite{THERMUS}), with plans to include other particles in the analysis, ultimately enabling a system-wide characterization of bulk particle production across the full STAR FXT energy range.

\noindent \textit{Acknowledgements.} This material is based upon work supported by the National Science Foundation under Grant No. 250721.
\bibliographystyle{elsarticle-num-notitle}
\bibliography{proceedings}
\end{document}